\documentclass[aps,prx,twocolumn,superscriptaddress, nofootinbib,  longbibliography]{revtex4-2}
\usepackage{graphicx} % needed for figs
\usepackage{dcolumn}  % needed for some tables
\usepackage{bm}    % for math
\usepackage{amssymb}  % for math
\usepackage{amsfonts, amsmath, amssymb, mathtools}
\usepackage{amsmath}

\usepackage{lipsum}
\usepackage[normalem]{ulem}
\usepackage{commath}
\usepackage{xcolor}
\usepackage{hyperref}

\usepackage{scalerel}[2016-12-29]

\usepackage[binary-units=true]{siunitx}
\usepackage{hyperref}
\usepackage{cleveref}
\RequirePackage{textcase}
\DeclareUnicodeCharacter{2009}{\,} 

\begin{document}

\newcommand{\condProb}{\operatorname{P}\expectarg}
\DeclarePairedDelimiterX{\expectarg}[1]{(}{)}{%
 \ifnum\currentgrouptype=16 \else\begingroup\fi
 \activatebar#1
 \ifnum\currentgrouptype=16 \else\endgroup\fi
}

\newcommand{\innermid}{\nonscript\;\delimsize\vert\nonscript\;}
\newcommand{\activatebar}{%
 \begingroup\lccode`\~=`\|
 \lowercase{\endgroup\let~}\innermid
 \mathcode`|=\string"8000
}

\newcommand{\todo}[1]{{\color[rgb]{1.0,0.0,0.0}#1}} % gaps to be filled in red
\newcommand{\err}{\epsilon} % Error generic
\newcommand{\bra}[1]{\left<#1\right|}
\newcommand{\ket}[1]{\left|#1\right>}
\newcommand{\bket}[2]{\left<#1~|~#2\right>}
\newcommand{\tr}[1]{\text{Tr}\left(#1\right)}
\newcommand{\kket}[1]{\left|\left|#1\right>\right>}
\newcommand{\bbra}[1]{\left<\left<#1\right|\right|}
\newcommand{\ba}{\boldsymbol{a}}
\newcommand{\bb}{\boldsymbol{b}}
\newcommand{\bc}{\boldsymbol{c}}
\newcommand{\bd}{\boldsymbol{d}}
\newcommand{\bh}{\boldsymbol{h}}
\newcommand{\bq}{\boldsymbol{q}}
\newcommand{\bp}{\boldsymbol{p}}
\newcommand{\bQ}{\boldsymbol{Q}}
\newcommand{\bP}{\boldsymbol{P}}
\newcommand{\bE}{\boldsymbol{E}}
\newcommand{\mE}{\mathcal{E}}
\newcommand{\Tr}{\text{Tr}}
\renewcommand{\Re}{\text{Re}}
\renewcommand{\Im}{\text{Im}}
\newcommand{\ta}{\tilde{\alpha}}
\newcommand{\bO}{\boldsymbol{\mathcal{O}}}
\newcommand{\br}{\boldsymbol{r}}
\newcommand{\bR}{\boldsymbol{R}}
\newcommand{\bK}{\boldsymbol{K}}
\newcommand{\bJ}{\boldsymbol{J}}
\newcommand{\bH}{\boldsymbol{H}}
\newcommand{\bU}{\boldsymbol{U}}
\newcommand{\bM}{\boldsymbol{M}}
\newcommand{\bX}{\boldsymbol{X}}
\newcommand{\bZ}{\boldsymbol{Z}}
\newcommand{\bY}{\boldsymbol{Y}}
\newcommand{\bI}{\boldsymbol{I}}
\newcommand{\bL}{\boldsymbol{L}}
\newcommand{\bT}{\boldsymbol{T}}
\newcommand{\bD}{\boldsymbol{D}}
\newcommand{\bn}{\boldsymbol{n}}
\newcommand{\bS}{\boldsymbol{S}}
\newcommand{\bsigma}{\boldsymbol{\sigma}}
\newcommand{\bSigma}{\boldsymbol{\Sigma}}
\newcommand{\bDelta}{\boldsymbol{\Delta}}
\newcommand{\bPi}{\boldsymbol{\Pi}}
\newcommand{\red}[1]{\textcolor{red}{#1}}
\newcommand{\green}[1]{\textcolor{green}{#1}}
\newcommand{\blue}[1]{\textcolor{blue}{#1}}
\newcommand{\bphi}{\boldsymbol{\varphi}}
\newcommand{\NN}{\mathcal N}

\raggedbottom

\renewcommand{\thesubsection}{\thesection.\arabic{subsection}}
\renewcommand{\thesubsubsection}{\thesubsection.\arabic{subsubsection}}

% \title{Protecting the quantum interference of cat states by phase-space compression
% }
\title{Shaping photons: quantum information processing with bosonic cQED}

\author{Adrian Copetudo}
\thanks{These authors contributed equally to this article. The order of author names can be re-arranged in individual CVs.}
\author{Clara Yun Fontaine}
\thanks{These authors contributed equally to this article. The order of author names can be re-arranged in individual CVs.}
\author{Fernando Valadares}
\thanks{These authors contributed equally to this article. The order of author names can be re-arranged in individual CVs.}
\affiliation{Centre for Quantum Technologies, National University of Singapore, Singapore}
\author{Yvonne Y. Gao}
\email[Corresponding author: ]{yvonne.gao@nus.edu.sg}
\affiliation{Centre for Quantum Technologies, National University of Singapore, Singapore}
\affiliation{Department of Physics, National University of Singapore, Singapore}
\date{\today}

\begin{abstract}
With its rich dynamics, the quantum harmonic oscillator is an innate platform for understanding real-world quantum systems, and could even excel as the heart of a quantum computer. A particularly promising and rapidly advancing platform that harnesses quantum harmonic oscillators for information processing is the bosonic circuit quantum electrodynamics (cQED) system. In this article, we provide perspectives on the progress, challenges, and future directions in building a bosonic cQED quantum computer. We describe the main hardware building blocks and how they facilitate quantum error correction, metrology, and simulation. We conclude with our views of the key challenges that lie on the horizon, as well as scientific and cultural strategies for overcoming them and building a practical quantum computer with bosonic cQED hardware.
\end{abstract}

\maketitle

The harmonic oscillator is a foundational model for physical systems. It is simple to understand and work with, yet has a spectrum of applications across classical and quantum physics. From everyday curiosities like oscillating guitar strings and pendulum clocks, to scientific endeavors with molecular vibrations and optical properties of matter, the harmonic oscillator arises in physics at all scales.
    
In the regime where quantum physics becomes relevant, the harmonic oscillator becomes a playground of quantum phenomena. With its versatile dynamics, the quantum harmonic oscillator (QHO) is an innate platform for encapsulating and understanding real-world quantum systems, and can even excel as the heart of a quantum computer. While a universal, fault-tolerant quantum computer still lives in the future, its potential innovations in cryptography~\cite{shor_polynomial-time_1997}, drug discovery~\cite{cao_potential_2018}, renewable energy~\cite{ajagekar_quantum_2022, giani_quantum_2021}, finance~\cite{orus_quantum_2019}, and more are propelling the efforts toward its realization.

What does it mean to compute with a QHO? It begins with understanding its dynamics. The QHO is conveniently described using the continuous variables (CV) of position $X$ and momentum $P$, and the typical operations on this system are expressed in a simple analytical form since they come from Hamiltonians linear or quadratic in $X$ and $P$~\cite{lloyd_quantum_1999}. These elementary operations, such as displacements, rotations, and squeezing, are also known as Gaussian operations. Within the Gaussian regime, several notable applications have been 
demonstrated, including quantum teleportation~\cite{braunstein_teleportation_1998, ralph_teleportation_1998, vaidman_teleportation_1994, bowen_experimental_2003, furusawa_unconditional_1998, zhang_quantum_2003}, cloning~\cite{cerf_cloning_2000, lindblad_cloning_2000}, communication with quantum dense coding~\cite{ban_quantum_1999, braunstein_dense_2000, ralph_unconditional_2002, mizuno_experimental_2005, pereira_quantum_2000}, and cryptography~\cite{hillery_quantum_2000, ralph_continuous_1999,reid_quantum_2000,cerf_quantum_2001, grosshans_continuous_2002, grosshans_quantum_2003}.

However, universal control of quantum computers cannot be achieved with just Gaussian operations~\cite{lloyd_quantum_1999}. Supplementing them with non-Gaussian operations (that come from higher-order Hamiltonians on X and P) enables the construction of arbitrary transformations necessary for quantum computational advantage~\cite{bartlett_efficient_2002, niset_no-go_2009}, entanglement distillation~\cite{eisert_distilling_2002}, and Bell inequality violations~\cite{bell_epr_2004}. To access non-Gaussian operations, and thus enable universal control, the QHO can be coupled to a single nonlinear element~\cite{lloyd_quantum_1999}. 

A system that facilitates both Gaussian and non-Gaussian operations can be realized in a variety of physical platforms. There is active research on the motional degree of freedom of trapped ions~\cite{bruzewicz_trapped-ion_2019}, electromagnetic fields of resonators~\cite{deleglise_reconstruction_2008,hacker_deterministic_2019}, Rydberg atom arrays~\cite{ebadi_quantum_2021}, flying photons~\cite{ourjoumtsev_generating_2006}, mechanical resonators~\cite{kotler_direct_2021, chu_perspective_2020} and other physical systems~\cite{hou_generation_2016, bulutay_cat-state_2017}. Different platforms present different advantages and challenges. A particularly promising and rapidly advancing candidate is bosonic circuit quantum electrodynamics (cQED), a platform that uses superconducting circuits.

\begin{figure*}
\centering
\includegraphics[width=1\textwidth]{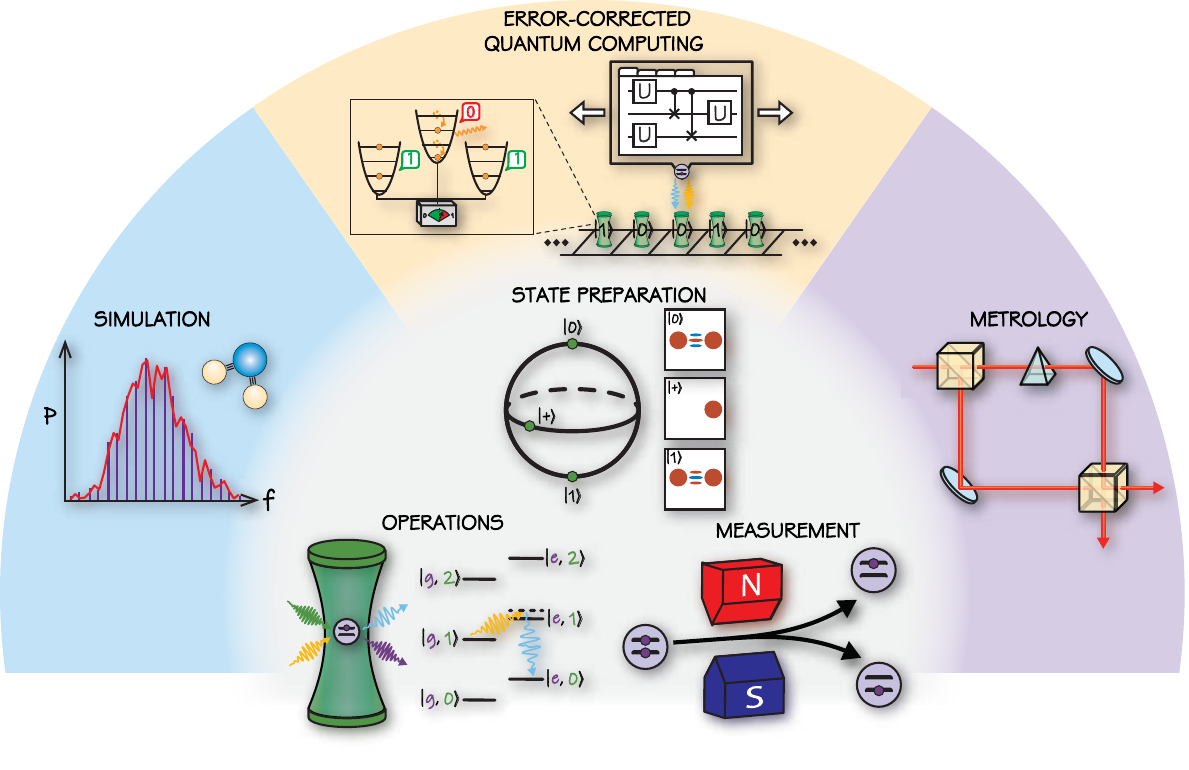}
\caption{Every quantum protocol involves preparing an initial state with the encoded information, processing it via gates or operations, and measuring the final state to gain knowledge about the computation result. These building blocks have been demonstrated in several tailored applications, including quantum metrology, quantum simulation, and error-correcting codes for quantum computing. In the future, we expect that quantum systems will be capable of supporting a much broader range of applications.}
%\caption{Every quantum computation involves preparing an initial state with the encoded information, processing it via gates or operations, and measuring the final state to gain knowledge about the computation result. These building blocks have been successfully demonstrated for several tailored applications, including quantum error correction, quantum metrology, and quantum simulation. Ideally, a universal fault-tolerant quantum computer will someday be capable of supporting a much broader range of applications.}
\label{fig: Fig1}
\end{figure*}

This article provides our perspective on the progress, challenges, and future directions in building a bosonic cQED quantum computer. We start with the description of the elementary hardware and describe how it facilitates state preparation, operations, and measurements. We then move into the more sophisticated tools and applications of error correction, metrology, and simulation. Finally, we conclude with our view of the key challenges that lie on the horizon, as well as scientific and cultural strategies towards overcoming them and building a universal, fault-tolerant quantum computer in bosonic cQED (Fig.~\ref{fig: Fig1}).

Bosonic cQED shines as a robust, accessible, and versatile platform for CV quantum experiments. QHOs can be readily realized with superconducting resonators, which are essentially LC circuits with well-defined resonance frequencies. The quantum information is stored in the electromagnetic field of these resonators, and it is prepared, transformed, and measured via the engineered interaction with a nonlinear auxiliary circuit. 

A good resonator stores information for at least as long as the computational time. In practice, experimental realizations of resonators are inherently imperfect: stored information decays over time through defects in the superconducting materials and their interfaces, and undesired couplings to the environment. Naïvely, the resonator field could be isolated from all external influences. But to control the state, the resonator must be coupled to an auxiliary circuit and the environment, unavoidably introducing more loss channels. This highlights a trade-off between protection and controllability and sets the strategy for engineering resonator geometries and materials. 

The most traditional choice for resonator architecture is the planar geometry, such as striplines and coplanar waveguides. They consist of superconducting metal circuits deposited on a dielectric substrate and can be fabricated with standard lithographic techniques that enable geometric precision and mass production. Coupling to auxiliary circuits is as simple as fabricating them on the same chip. The earliest on-chip resonators, however, only achieved lifetimes of tens of microseconds due to the high participation of the electromagnetic fields in the lossy substrate bulk and interfaces~\cite{calusine_analysis_2018, megrant_planar_2012}. These limited lifetimes inspired the exploration of 3D resonators. With 3D resonators, long lifetimes on the order of milliseconds are achieved reliably with a simple cylindrical or rectangular superconducting resonator~\cite{reagor_reaching_2013}. With nearly all of the field in vacuum, the dominant loss only acts on the fringes of the field that interact on the surface of the superconductor. The TESLA geometry was optimized with this in mind and has the lowest relative field at the superconductor surface across 3D resonator geometries~\cite{aune_superconducting_2000}. 

However, designing solely to minimize the field on the resonator surface impedes the ability to couple to an external auxiliary circuit, calling for more creative designs. One approach is to introduce a center pin in the cylindrical cavity, transforming it into a $\lambda/4$ coaxial resonator~\cite{reagor_quantum_2016}. With the field now densely concentrated at the pin's termination, the auxiliary circuit can be housed in a coaxline architecture~\cite{axline_architecture_2016} adjacent to the resonator. This higher degree of isolation can achieve strong coupling while still preserving long resonator lifetimes of up to a few milliseconds~\cite{reinhold_error-corrected_2020, kudra_high_2020, milul_superconducting_2023}.

Beyond optimizing the protection and controllability of a single memory, it is valuable to develop an architecture that can scale to multiple memories. One approach for this scale-up comes from reducing the physical dimensions of the resonators, making the devices more compact. This is the case for micromachined 2.5D resonators~\cite{brecht_multilayer_2016, brecht_micromachined_2017}, which benefit from on-chip fabrication processes while leveraging the higher lifetimes of 3D architectures. Another approach is to utilize several electromagnetic field configurations, each acting as an independent memory, within a single resonator. This is realized in the 3D flute resonator~\cite{chakram_seamless_2021}, a meandering 3D tunnel that is low-frequency, seamless, and simple to manufacture, as well as in the double-post resonator~\cite{koottandavida2023erasure}.

For each geometry, further improvements to lifetimes come from improving the quality of lossy elements in the device. For instance, seam loss from metal contact resistance can be reduced with annealing, electron-beam welding, and indium bonding~\cite{brecht_demonstration_2015}. An active area of investigation is reducing surface and material losses that arise from the superconductor and dielectrics. Every material has loss characteristics --- such as interface properties, compatibility with surface treatments, and quality of fabrication --- that present a plethora of considerations when choosing a material. For example, replacing high-purity aluminum with niobium with the same  $\lambda$/4 coaxial geometry resulted in higher lifetimes~\cite{heidler_non-markovian_2021}. In the case of on-chip architectures, tantalum resonators deposited on annealed, high-grade sapphire dielectric have surpassed the state-of-the-art aluminum resonators~\cite{ganjam_surpassing_2023, crowley_disentangling_2023}. However, the quantitative attribution of lifetime to specific loss mechanisms remains challenging, as losses may not be separable based on what can be measured and simulated~\cite{crowley_disentangling_2023, lei_characterization_2023}.

\begin{figure*}
\centering
\includegraphics[width=1.0\textwidth]{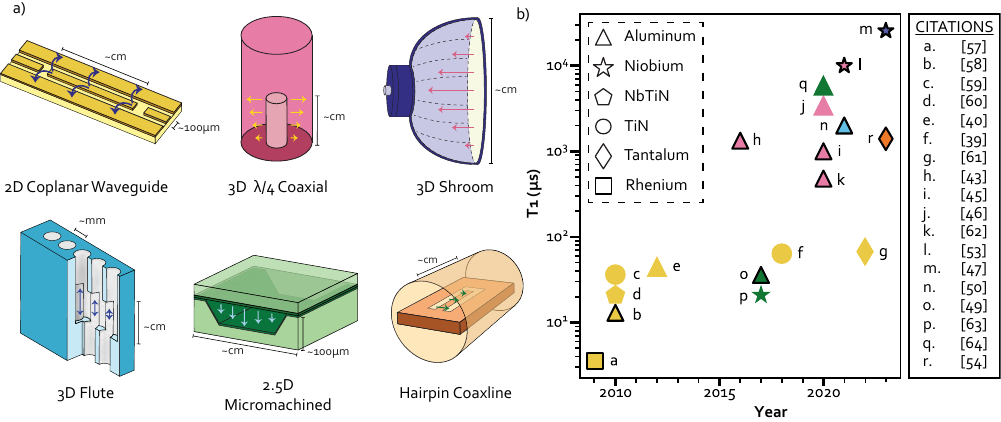}
\caption{Designs of superconducting resonators and their lifetimes. (a) Illustrations of six popular resonator designs in bosonic cQED, with indications of their approximate sizes. (b) A selected collection of resonator T$_1$ lifetimes for the six illustrated resonator designs extracted from the literature. The lifetimes of resonators have been steadily increasing over the last 15 years due to the engineering of materials and geometries. The shape of the points indicates the superconducting material, and the color corresponds to the adjacent diagrams. We highlight studies that successfully integrate one or more nonlinear circuits with a black border. References: 2D: \cite{hofheinz_synthesizing_2009, leek_cavity_2010, vissers_low_2010,barends_minimal_2010,megrant_planar_2012, calusine_analysis_2018, shi_tantalum_2022}, 3D coaxial: \cite{reagor_quantum_2016, reinhold_error-corrected_2020,kudra_high_2020, ma_error-transparent_2020, heidler_non-markovian_2021}, 3D shroom: \cite{milul_superconducting_2023}, 3D flute: \cite{chakram_seamless_2021}, 2.5D micromachined: \cite{brecht_micromachined_2017,zoepfl_characterization_2017,lei_high_2020}, hairpin coaxline: \cite{ganjam_surpassing_2023}}.
\label{fig: Fig2}
\end{figure*}

Through concurrent enhancements of geometries and materials driven by understandings of loss mechanisms, there has been meaningful growth in resonator lifetimes (Figure~\ref{fig: Fig2}). Recently, the niobium shroom resonator --- an adaptation of the TESLA design --- has demonstrated both successful integration with an auxiliary circuit and a record lifetime of 34ms~\cite{milul_superconducting_2023}. In a more scalable approach, lifetimes surpassing one millisecond were achieved with tantalum resonators in hairpin-shaped coaxline geometry~\cite{ganjam_surpassing_2023}. With this result, we believe on-chip architectures could emerge as the leading memory design in bosonic cQED with their inherent scalability and ease of production.

Apart from resonator geometry and material, the quality of the coupled auxiliary circuit for control and measurement also impacts the fidelity of storage of information and operations. Auxiliary circuits could hamper resonator lifetimes by introducing dielectric losses or through the Purcell effect. They also can introduce unwanted non-Gaussian distortions to the resonator dynamics. In developing an auxiliary circuit with high-performing control, these limiting factors must be within sight.

All nonlinear circuits in cQED are based on the Josephson junction, a nano-fabricated element of two superconducting islands bridged by an oxide tunnel barrier. Its behavior can be modeled as a nonlinear inductor. When the capacitance of the two islands is large enough, the circuit becomes insensitive to charge noise~\cite{koch_charge-insensitive_2007}. Popularly referred to as a transmon, this circuit behaves as an anharmonic resonator due to its 4th-order nonlinearity in $X$ and $P$. This nonlinearity can effectively limit operations to the first two energy levels of the transmon, making it a prevailing superconducting qubit for discrete-variable quantum processors~\cite{krinner_realizing_2022, arute_quantum_2019, kim_evidence_2023}. 

For CV quantum information processing, it is more interesting to investigate the non-Gaussian operations that the transmon anharmonicity grants to the resonator. The superconducting islands of the transmon can be extended to overlap with the electromagnetic field of the resonator, coupling both circuits together. The strength of this interaction can be engineered by adjusting the detuning between their resonance frequencies or changing the overlap of their electromagnetic fields. Typically the circuits are designed to operate in the dispersive coupling regime, where the detuning between the transmon and the resonator is much larger than their coupling. In this regime, the dynamics of the resonator and the transmon become interdependent: the resonance frequency of one element shifts depending on the energy state of the other.

The coupled dynamics allow complete characterization of the state of the resonator. Properties of the resonator state across phase space can be mapped onto the transmon, where they are extracted by direct measurements. The simplest properties to measure with the dispersive shift are photon number parity with a standard Ramsey technique~\cite{lutterbach_method_1997}, and photon number distribution with frequency-selective transmon pulses~\cite{schuster_resolving_2007}. Current techniques to extract figures of interest require redundant and time-consuming measurements. We expect more efficient strategies to emerge in the future, such as quantum reservoir processing~\cite{ghosh_quantum_2019} and shadow tomography~\cite{huang_predicting_2020}.

Complementarily, the dispersive interaction is harnessed to prepare and control arbitrary resonator states through a variety of techniques, such as SNAP gates~\cite{krastanov_universal_2015}, echo-conditional displacement (ECD) gates~\cite{eickbusch_fast_2022}, and numerically-optimized drive pulses resonant with the transmon and the resonator~\cite{heeres_implementing_2017}. Furthermore, the transmon anharmonicity offers other tools for state manipulation that can be activated on demand with off-resonant drives, such as the Stark shift~\cite{zhang_engineering_2019}, two-photon pumps~\cite{leghtas_confining_2015}, photon blockade~\cite{bretheau_quantum_2015}, and beamsplitting between oscillators~\cite{gao_programmable_2018}. 

Although the transmon equips us with numerous tools for manipulation and measurement, it can also introduce undesirable effects to the resonator. One such effect comes from the dielectric loss of the Josephson junction itself. The participation of the resonator field is significant in this region, so the losses at the junction directly impact resonator lifetimes~\cite{wang_surface_2015}. The junction is also sensitive to high-energy radiation~\cite{jin_thermal_2015}, which can induce excitation of the transmon and thus impact the fidelity of operations and measurements. Other known losses incurred by the junction also include quasiparticles and spurious two-level systems~\cite{spiecker_two-level_2023}.

In addition to these incoherent errors, the higher harmonics of the junction can also impart unwanted dynamics of the resonator~\cite{willsch_observation_2023}. Furthermore, the resonator --- ideally harmonic --- inherits a small amount of anharmonicity called self-Kerr that is always present and cannot be trivially averted. Thus, it can be a limiting factor to the fidelity of memory operations. One way to reduce self-Kerr is to decrease the coupling strength, but this comes at the expense of slower dispersive interactions. It was recently shown that fast operations can still be realized even in this weak coupling regime with ECD gates~\cite{eickbusch_fast_2022}, such as compression of a cat state~\cite{pan_protecting_2023} and creation of binomial and Gottesman-Kitaev-Preskill (GKP) logical states for error correction~\cite{eickbusch_fast_2022}. Conveniently, this weaker coupling regime dampens the inherited decay mechanisms from the transmon, bolstering the ECD control scheme to be effective in experiments highly dependent on a robust memory lifetime~\cite{sivak_real-time_2023}. 

Another approach is to explore circuits with different nonlinear behaviors and energy landscapes. For example, the transmon can be adapted by adding a second Josephson junction in parallel to create a superconducting quantum interference device (SQUID)~\cite{koch_charge-insensitive_2007}. The resonance frequency of the SQUID is sensitive to the flux that threads the loop formed by the two junctions, allowing for frequency tunability by an external magnetic field. By placing its frequency closer or further from the resonator frequency, the coupling can be switched on and off to minimize undesirable dynamics --- such as resonator self-Kerr --- during idle periods of an experiment. Following a similar thread, the superconducting nonlinear asymmetric inductive element (SNAIL)~\cite{frattini_3-wave_2017} is another flux-sensitive device formed by three large junctions in parallel to a smaller one. By tuning the magnetic field, its 4th-order nonlinearities can be reduced, eliminating self-Kerr. At the same time, the 3rd-order nonlinearities can be enhanced and used to control the resonator states via three-wave mixing~\cite{hillman_universal_2020, chapman_high--off-ratio_2023}. Many other alternative circuits under study~\cite{gyenis_moving_2021, rasmussen_superconducting_2021} differ in fabrication complexity, protection against loss, and their nonlinear dynamics. A common challenge in operating these circuits is integrating flux control with long-lived resonators. However, there are promising recent solutions to this limitation~\cite{valadares_ondemand_2023, gargiulo_fast_2021, lu_high-fidelity_2023, chapman_high--off-ratio_2023}, and future developments in this area will allow us to fully explore the variety of auxiliary circuits~\cite{groszkowski_scqubits_2021, miano_hamiltonian_2023}. 

There has been remarkable progress in the development of cQED hardware for a quantum computer. The essential building blocks are coming together as basic quantum processors in increasingly sophisticated ways. However, despite having universal control of quantum information, the hardware is inherently noisy and prone to error; the quantum information does not yet live long enough to be of practical use. Quantum error correction (QEC) thus stands as a core pillar for building a fault-tolerant quantum computer. 

A QEC protocol encodes the abstract unit of quantum information --- a logical qubit --- into distinguishable states of the system, called codewords. Any linear combination of these codewords defines the logical code space. These states must share a measurable observable that indicates when the resonator state moves from the code space to an error space. This error is either corrected in situ or accounted for by classical post-processing of the measurement outcomes. 

In bosonic cQED, the realization of a large phase space (equivalently, many energy levels) in a QHO offers hardware-efficient QEC schemes where one logical qubit is encoded in a single resonator. The errors appear as properties of the same resonator state and thus are detectable with local measurements.

One of the most noteworthy bosonic encodings is the GKP code~\cite{gottesman_encoding_2001}, which leverages the non-commutative geometry of phase space to protect against errors. The logical states, in theory, are infinite grids of infinitely-squeezed peaks in phase space. One logical state is transformed into the other with a displacement operation, making the code translationally symmetric. In reality, for finite energy considerations, GKP states are generated as a grid of squeezed peaks occupying a finite area in phase space. While the preparation of GKP states requires non-Gaussian operations, arbitrary quantum operations within the logical code space can be implemented with just Gaussian operations~\cite{gottesman_encoding_2001}. Both dominant loss mechanisms --- energy decay and dephasing --- cause random displacements of the peaks, leading to the diffusion of their representation in phase space. By frequently detecting and correcting for these loss-driven displacements, errors are detected and logical states are actively stabilized while performing computation. The suppression of logical errors and extension of the logical qubit lifetime using the GKP code was a recent experimental milestone~\cite{campagne-ibarcq_quantum_2020, sivak_real-time_2023}.

In an alternative class of codes, known as rotation-symmetric codes, errors manifest as parity flips. Binomial codes~\cite{michael_new_2016} use logical codewords that are superpositions of photon number states with the same parity. Upon a photon loss event, the parity of the codewords flips, signaling the occurrence of the error~\cite{hu_quantum_2019}. Cat codes use cat states, which are superpositions of coherent states. The simplest one that can correct for photon loss is the 4-component cat code~\cite{leghtas_hardware-efficient_2013}, where the codewords are the even-parity cat states along both axes in phase space. Similarly to the binomial code, the parity is used to identify photon loss. Both the cat and binomial codes have been demonstrated to surpass the break-even point~\cite{ofek_extending_2016, gertler_protecting_2021, ni_beating_2023}. 

While the GKP code can detect all dominant errors and stabilize the codewords with active feedback, robust implementation is challenging. On the other hand, cat and binomial codes are easier to implement and can detect one type of error effectively. They can be also stabilized through engineered dissipation and the use of multi-photon pumps~\cite{puri_bias-preserving_2020,grimm_stabilization_2020, gertler_protecting_2021, lescanne_exponential_2020, berdou_one_2023}. This hints that the road to fault-tolerant computation will most certainly require concatenating several complementary QEC layers; for instance, using cat states that protect against photon loss embedded in a surface code~\cite{kitaev_fault-tolerant_2003,chamberland_building_2022, regent_high-performance_2023} that corrects for dephasing and other less common sources of error. In parallel, active research is also underway to explore codewords that span over multiple resonators. These are more hardware-demanding in preparation but could potentially be more comprehensive in their QEC capacities~\cite{lu_high-fidelity_2023, chou_demonstrating_2023, teoh_dual-rail_2022, gertler_experimental_2023}.

Many feedback-based QEC techniques rely on performing high-precision measurements to promptly detect errors. Quantum metrology focuses on performing measurements of physical parameters at the fundamental limits of sensitivity and resolution imposed by quantum mechanics.\cite{sahota_quantum_2014}. Examples of these parameters are the phase of a quantum state, which can be used in gravitational wave detectors ~\cite{abadie_gravitational_2011} and magnetometers~\cite{danilin_quantum-enhanced_2018}, or the amplitude of a drive~\cite{wang_heisenberg-limited_2019}, which can be used to create very precise displacements or measure electric fields.

The usual method of conducting a precise measurement in the presence of noise is to perform $N$ measurements and average the results; the estimation uncertainty then scales at a rate of $1/\sqrt{N}$, which is known as the standard quantum limit (SQL). By making use of quantum resources, such as entanglement, squeezing, or non-Gaussianity, this limit can be surpassed and theoretically even reach the Heisenberg limit (HL), where the uncertainty decreases with $1/N$. 

The archetype of a metrology experiment is the Mach-Zehnder interferometer (MZI), a two-mode interferometer used for phase estimation. One way to realize the analog of the MZI in bosonic cQED is with two superconducting resonators, which can be thought of as the arms of the interferometer. The bosonic states stored in them undergo a beam-splitter transformation, a phase shift on one of the resonators, then a second beam-splitter operation, followed by a mean-photon number distribution readout of both resonators. With this setup, an on-demand phase shifter can be applied selectively on one of the modes within an experiment~\cite{gao_programmable_2018}.

In the case of a QHO, the SQL is determined by the uncertainty of coherent states in phase space. Hence, sub-SQL precision can be reached by using states whose phase space distributions have structures at smaller scales than the coherent state uncertainty~\cite{penasa_measurement_2016}, also called sub-Planck structures. Quantum resources such as entanglement or multimode squeezed states~\cite{sahota_quantum_2014} offer these sub-Planck structures and enable measurement precision at the HL. However, the coherence of such multimode states is particularly sensitive to losses, which could jeopardize the quality of the detection. 

Fortunately, a single resonator state suffices to beat the SQL and even reach the HL~\cite{wang_heisenberg-limited_2019}. This is particularly attractive because a single resonator is potentially less sensitive to non-local perturbations, hardware-efficient and easily engineered in bosonic cQED. Single-resonator states that feature sub-Planck structures, such as the 0N state~\cite{wang_heisenberg-limited_2019}, cat state~\cite{munro_weak-force_2002}, squeezed vacuum~\cite{sahota_quantum_2014}, compass states, and GKP states~\cite{duivenvoorden_single-mode_2017}, are valuable resource states that enable measurement precision at the HL.

Reaching higher levels of precision comes with the enhanced ability to fine-tune state preparation, operations, and measurements. These core capabilities also drive exciting developments in another powerful application in bosonic cQED: analog simulation. 

Analog simulators are tailored application-dependent devices that naturally undergo the same dynamics as the simulated system, such that there is a direct relation between the observables of the simulator and the simulated system~\cite{braumuller_analog_2017}. Bosonic cQED is an ideal platform for such simulators due to its capability to engineer tailored parameters and interactions. Since both bosons and resonators are QHOs, a 1-to-1 mapping is established between their energy levels without an intermediary translation. This extends to boson-to-boson and boson-to-fermion interactions, as they can be directly simulated by a resonator coupled to another resonator or a nonlinear auxiliary circuit, respectively. An example of this is the simulation of the absorption spectra of diatomic molecules~\cite{wang_efficient_2020, hu_simulation_2018}; the phononic energy levels of the nuclear vibrational motion of the molecule are encoded into the resonator, and the electronic structure of the molecule is encoded into the two levels of the coupled transmon. Through conventional state preparation, operations, and measurements, the electronic transition dipole correlation function~\cite{hu_simulation_2018} or the Franck-Condon factors~\cite{wang_efficient_2020} can be computed, which are then used to calculate the excitation spectrum of the molecule. These bosonic simulations can also be extended to study the quantum Rabi model~\cite{braumuller_analog_2017}, photon-photon interactions~\cite{houck_-chip_2012} or quantum magnetism~\cite{kurcz_hybrid_2014}, and conical intersection~\cite{wang_observation_2023}.

Over the past 10 years, the bosonic cQED platform has advanced at an impressive pace. Breakthroughs in coherence times, functionalities, and control and tomography protocols made this platform a versatile testbed for CV experiments.
Its applications to quantum error correction, metrology, and simulation of quantum systems are becoming increasingly reliable, pointing to a future where these functionalities merge as routines of a single quantum processor. We see no fundamental roadblocks on the horizon, giving us a vast space for sophisticated refinement and creative exploration. But there is still a long way to go before universal, fault-tolerant quantum computing becomes a reality. Toward this goal, we see these questions as essential: How to minimize loss through geometries, materials, and fabrication processes? How to introduce rich yet non-disruptive dynamics through engineered nonlinearities? How to extract information in an efficient and adaptable way? 

As the single bosonic logical qubit becomes more advanced, another pertinent area of investigation is the integration of multiple logical qubits together, and the realization of a universal set of gates~\cite{gao_programmable_2018, rosenblum_cnot_2018}. To enable this transition from small prototypical devices to large-scale quantum computing, we must tackle the challenge of scalability. Thus, technological advancements in superconducting circuits themselves, cryogenic equipment, control electronics, and so on, will play a significant role.

Fruitful explorations of these research directions will rely on the cross-pollination of ideas and techniques across disciplines. The cooperation between different fields has already been very successful: quantum optics laid the theoretical groundwork of Gaussian quantum information; microwave engineering techniques are readily employable to build low-noise cQED control systems; a more nuanced understanding of material structures and defects boosts coherence times for resonators and transmons; converging interests across quantum imaging, dark matter detection, high-precision quantum state preparation, etc. is accelerating the development of a microwave single-photon detector~\cite{lamoreaux_analysis_2013, brubaker_first_2017}; the theory of quantum error correction can be applied to study black holes~\cite{preskill_physics_2022}; and so on. These synergistic collaborations across disciplines have been and will continue to be essential for scientific progress. 

Beyond discipline, harnessing diversity in our scientific community is vital in creating meaningful and far-reaching quantum technologies. To do so, we must first recognize that the impact felt from a technology inevitably depends on the individual's unique interactions with the surrounding social, economic, and environmental systems ~\cite{buolamwini_gender_2018, beaunoyer_covid-19_2020, johnson_intersectionality_2020}. These systems are historically entangled, interconnected, and ever-changing~\cite{crenshaw_demarginalizing_1989}, leading to profound implications: quantum technology may not just provide solutions; it may also generate and perpetuate inequalities~\cite{zheng_inequality_2021, de_wolf_potential_2017,ten_holter_reading_2023, divincenzo_scientists_2017}. The landscape of these impacts is intertwined with the way we construct knowledge and technology. To secure a future of quantum computing that is attuned to the people it will serve, we must transform our scientific culture to embrace diverse perspectives, provide equitable access to use and contribution, and foster a sense of belonging and empowerment to each participant. Finding power in difference not only enriches scientific innovation~\cite{freeman_collaborating_2015, alshebli_preeminence_2018, hofstra_diversityinnovation_2020, seskir_democratization_2023} but also allows quantum computing to benefit people across identities and experiences.

Building a universal quantum computer is a multidisciplinary pursuit. Bosonic cQED is a rapidly growing area that orchestrates a diverse community of people and disciplines. In the coming decade, we envision it to become increasingly mature and sophisticated, while continuously re-inventing itself. With its innate versatility, bosonic cQED can also integrate with other platforms, leading to the emergence of future research directions. But one thing is for certain, bosonic cQED is here to stay and will be an integral part in shaping the future of quantum science and technology. 

\section*{Acknowledgment}
We thank Dr. Luigi Frunzio, Prof. Gerhard Kirchmair, and Prof. Ioan Pop for their valuable comments and feedback on the article. Y.Y.G. acknowledges the funding support of the National Research Foundation Fellowship (NRFF12-2020-0063) and the Ministry of Education, Singapore.

\section*{Data Availability}
Data sharing is not applicable to this article as no new data were created or analyzed in this study.

%\bibliographystyle{unsrt}
%\bibliographystyle{apsrev4-1}

% \bibliography{references}    %use a bibtex bibliography file


\end{document}